\documentclass[%
 reprint,
 showkeys,
 amsmath,amssymb,
 aps,
prc]{revtex4-2}

\usepackage{graphicx}
\usepackage{color}
\usepackage[normalem]{ulem}
\usepackage{dcolumn}
\usepackage{bm}
\usepackage[dvipsnames]{xcolor}

\begin{document}

\preprint{APS/123-QED}

\title{Pauli Blocking effects in Nilsson states of weakly bound exotic nuclei
} 


\author{P. Punta}
 \email{ppunta@us.es}
 \affiliation{%
 Departamento de FAMN, Facultad de Física, Universidad de Sevilla, Apartado 1065, E-41080 Sevilla, Spain
}%
\author{J. A. Lay}
\author{A. M. Moro}
\affiliation{%
 Departamento de FAMN, Facultad de Física, Universidad de Sevilla, Apartado 1065, E-41080 Sevilla, Spain
}%
\author{G. Col\`o}
\affiliation{%
 Dipartimento di Fisica, Universit\`a degli Studi di Milano, via Celoria 16, I-20133 Milano, Italy
}%
\affiliation{INFN, Sezione di Milano, via Celoria 16, I-20133 Milano, Italy}

\date{\today}

\begin{abstract}
\begin{description}
\item[Background] The description of weakly bound nuclei using deformed few-body models has proven to be crucial in the study of reactions involving certain exotic nuclei.
However, these core+valence models face the challenge of applying the Pauli exclusion principle,
since the factorisation of the system does not allow complete antisymmetrization.
Therefore, states occupied by core nucleons should be blocked for the valence nucleons.
\item[Purpose] We aim to study the Carbon isotopes $^{17}$C and $^{19}$C which are good examples of weakly bound exotic nuclei with significant deformation where the valence shell is partially filled.
We will explore the effect of different methods of blocking occupied states in deformed two-body models.
\item[Methods] 
The structure of $^{17}$C and $^{19}$C is described with deformed two-body models where a Nilsson Hamiltonian is constructed using Antisymmetrized Molecular Dynamic calculations of the cores.
Different methods of blocking occupied Nilsson states are considered using the Bardeen–Cooper–Schrieffer formalism: without blocking, total blocking and partial blocking.
The latter also takes into account pair correlations to some extent.
These models are later used to study $^{16}\text{C}(d,p)^{17}\text{C}$, $^{17}\text{C}(p,d)^{16}\text{C}$ and $^{18}\text{C}(d,p)^{19}\text{C}$ transfer reactions within the Adiabatic Distorted Wave Approximation. 
In the first case, the results are compared with experimental data.
\item[Results]
A good reproduction of the structure of $^{17}$C is found, significantly improving the agreement in the $^{16}\text{C}(d,p)^{17}\text{C}$ reaction including blocking effects. 
The $^{19}$C spectrum is better reproduced considering blocking, in particular, the partial blocking method that considers the pairing interaction provides the best description.
\item[Conclusions] Promising results are shown for the study of transfer reactions involving weakly bound exotic nuclei, by highlighting the effect of blocking occupied Nilsson states.
We envision to extend the models to the study of breakup reactions and to newly discovered halo nuclei.
\end{description}
\end{abstract}

\maketitle


\section{\label{sec:level1}Introduction}

In recent years, the study of weakly bound exotic nuclei has been boosted by the development of radioactive beam facilities.
The exotic nuclei have a rather different ratio of protons to neutrons from that of stable nuclei.
Because of that, their properties can be very different,
for example, some of them exhibit a halo nature.
Halo nuclei are weakly bound systems composed of a relatively compact core and 
one or two highly delocalized valence particle(s) forming a \textit{halo} of matter around the core.

To study the reactions including weakly bound nuclei,
they have usually been described using few-body models that ignore fragment deformations.
However, it is known that \textit{core deformations} can significantly affect both the structure and the dynamics of these systems{~\cite{Mor12,Mor12a,Del13,Lay16}}.
Therefore, to obtain a more reliable description of reactions involving these nuclei,
deformation needs to be included in few-body models.

Nuclei composed of a weakly bound neutron and an even-even core with significant quadrupole deformation are considered.
Different deformed two-body models have been successfully applied to the description of $^{17}$C, $^{19}$C and $^{11}$Be.
Promising results are obtained with the semi-microscopic PAMD model \cite{Lay14}, in which the Hamiltonian is constructed using the transition densities of the corresponding cores calculated in the Antisymmetrized Molecular Dynamics (AMD) formalism.
However, in the case of $^{17}$C, better results are obtained with the Nilsson model~\cite{Punta23}. 
In the present work, both models are combined in order to have a model like Nilsson's but maintaining most of the microscopic information of the PAMD model.

A inherent difficulty in deformed two-body models is the application of the Pauli exclusion principle related to the states occupied by the core nucleons.
In many-body calculations,
this problem is automatically solved by antisymmetrization of the wavefunction,
while few-body models require the explicit exclusion of forbidden states.
The simplest way to deal with this problem is to discard the bound states that are considered occupied by comparison with the spherical and Nilsson limits.
However, despite its widespread application,
this method shows limitations in its application to some nuclei.
Total blocking of single-particle states emerges as an alternative method,
initially proposed for spherical single-particle states and recently extended to deformed Nilsson states~\cite{Shin24}.

In this work, the possibility of including more sophisticated Pauli blocking effects is explored.
In particular, single-particle Nilsson states are totally or partially blocked using the Bardeen–Cooper–Schrieffer (BCS) formalism, thus including a residual pairing interaction.
In this way, pair correlations are also approximately taken into account in order to obtain more complete and descriptive models. 
The implementation of BCS theory in a deformed two-body model has already been tested~\cite{Tar06}.
However, in this new method,
instead of applying the BCS calculation to the spherical single-particle levels,
we use the deformed Nilsson single-particle levels as a starting point.

We present the results of using different Pauli blocking methods applied to the deformed weakly-bound nuclei $^{17}$C and $^{19}$C. 
The energies of the bound states and some low-lying resonances, together with their associated wavefunctions, are obtained by diagonalizing the model Hamiltonian in a basis of square-integrable functions.
We use the transformed harmonic oscillator functions (THO)~\cite{kar05}, which have been successfully applied to the discretization of the continuum of weakly bound nuclei for its application to breakup and transfer reactions both for two-body and three-body systems. \cite{Mor12a,Lay14,Punta23,Lay10,Lay12,manoli08,Casal16}. 

The calculated $^{17}$C wavefunctions are tested by applying them to the transfer reaction $^{16}\text{C}(d,p)^{17}\text{C}$, comparing with the experimental data from GANIL~\cite{Pereira}.
We will also propose similar reactions not yet experimentally measured, $^{17}\text{C}(p,d)^{16}\text{C}$ and $^{18}\text{C}(d,p)^{19}\text{C}$, and analyse how they can be used to discriminate between the different models presented.  

The paper is organized as follows. Sec.~II. describes the structure formalism. 
It focuses on the description of the new NAMD model and the implementation of the BCS theory.
Section III shows the results of the application to $^{17}$C, 
including the study of transfer reactions $^{16}\text{C}(d,p)^{17}\text{C}$ and $^{17}\text{C}(p,d)^{16}\text{C}$.
The application to $^{19}$C and $^{18}\text{C}(d,p)^{19}\text{C}$ are presented in Sec.~IV.
Finally, in Sec.~VI we summarize the main results of this work.

\section{\label{sec:Structure}Structure Formalism}

We consider systems that can be described using two-body models, where a neutron moves in a deformed potential generated by the core.
The Hamiltonian of the system can be written as 
\begin{equation}\label{eq:H}
{\cal H}=T(\vec r)+V_{\ell s}(r)(\vec \ell\cdot\vec s)+V_{vc}(\vec r,\xi)+h_{core}(\xi),
\end{equation}
where $T(\vec r)$ is the kinetic energy operator for the relative motion between the valence and the core and 
$h_{core}(\xi)$ is the Hamiltonian of the core.
$V_{vc}(\vec r,\xi)$ is the effective valence-core interaction,
which depends on the relative motion between the valence and the core,
but also on the core degrees of freedom $\xi$.
A spin-obit term with the usual radial dependence $V_{\ell s}(r)$
is added to this valence-core interaction.

The eigenvalues and eigenfunctions of the Hamiltonian are obtained by diagonalization in the THO basis.
In our case, the single-particle Nilsson Hamiltonian $\mathcal H_N=\mathcal{H}-h_{core}$ is diagonalized in the intrinsic system which rotates jointly with the core.
Later, the wave functions are projected in the fixed laboratory frame to finally add the term $h_{core}$. A more complete description of this method can be found in Sec.~II of Ref.~\cite{Punta23}.
The eigenfunction associated with an eigenvalue $\varepsilon^{J^\pi}_i$ can be generically expressed as
\begin{equation}\label{eq:LabWF}
\Psi_{iM}^{J^\pi}(\vec{r},\xi)=\sum_{\alpha}
R^{J^\pi}_{i\alpha}(r)\Phi_{\alpha J}^{M}(\hat{r},\xi).
\end{equation}
They are characterized by the index $i$, parity $\pi$,
the total angular momentum $J$ and its projection $M$ on the $z$~ axis of the fixed laboratory frame.
The radial functions $R^{J^\pi}_{i\alpha}(r)$ are a linear combination of orthonormal functions of the basis $R^{THO}_{n\ell}(r)$.
$\Phi_{\alpha J}^M(\hat{r},\xi)$ are the eigenfunctions of $J^2$ and $J_z$ 
resulting from the coupling of the angular momentum $\vec{j}$
of the valence particle to the core angular momentum $\vec{I}$,
\begin{equation}\label{eq:LabAng}
\Phi_{\alpha J}^M(\hat{r},\xi)\equiv\left[  
{\cal Y}_{\ell s}^j(\hat{r}) \otimes \phi_{I}(\xi) \right]_{JM}.
\end{equation}
Here, $\ell$ is the orbital
angular momentum of the valence particle relative to the core,
which couples to the spin of the valence particle $s$ to give the
particle total angular momentum $j$.
The label $\alpha$ denotes the set of quantum numbers $\{\ell,s,j,I\}$.
${\cal Y}_{\ell s}^{jm}(\hat{r})$ denotes the wavefunction resulting
from coupling the spin of the valence particle with the corresponding spherical harmonic.

The weight of each $\alpha$ component is given by the integral $\int dr|rR^{J^\pi}_{i\alpha}(r)|^2$.
When antisymmetrization between the valence neutron and the core is properly taken into account,
these weights correspond to spectroscopy factors (SF).
In our case, where antisymmetrization between the valence particle and the remaining nucleons is not considered, this correspondence between weights and SF is only approximate.
\subsection{\label{sec:NAMD}NAMD model}
In the method proposed here, denoted  NAMD hereafter, we combine the Nilsson and PAMD models compared in Ref.~\cite{Punta23}.
On the one hand, as in the Nilsson model,
the valence-core interaction is considered in the intrinsic frame ($\vec{r'}$),
which rotates jointly with the core.
For this frame, considering an axially symmetric quadrupole deformation,
we can assume that the potential $V_{vc}$ only depends on the radial coordinate $r=r'$ and
the angle $\theta'$ with respect to the symmetry axis of the core:
\begin{equation}\label{eq:NilssonPot}
 V_{vc}(r,\theta')=
 V_0(r)Y_{00}+V_2(r)Y_{20}(\theta').
\end{equation}
Furthermore, as in the Nilsson model, the core is approximated by a perfect rotor.
Therefore, $h_{core}$ depends on the angular momentum of the core $\vec I$ and its moment of inertia $\mathcal J$,
\begin{equation}
 h_{core}=
 \frac{\hbar^2}{2\mathcal J}\vec I^2.
\end{equation}

On the other hand, the coupling potentials $V_0(r)$ and $V_2(r)$ are calculated as in the PAMD model :
\begin{equation}
V_\lambda(r)=\langle0^+||V_\lambda(\vec r,\xi)||\lambda^+\rangle.
\end{equation}\label{eq:Vlambda}
The calculation of these reduced matrix elements follows the procedure of Ref.~\cite{Lay14}, where the nucleon-nucleon interaction of Jeukenne, Lejeune, and Mahaux~\cite{JLM} is convoluted
with microscopic transition densities of the core nucleus calculated with Antysymmetrized Molecular Dynamics (AMD) \cite{Kan13b}. 
In order to compare the strength of these couplings with a standard particle-rotor or Nilsson models, a  deformation length $\delta_2 \equiv\langle 0||\hat\delta_2||2\rangle$ can be extracted from the same transition densities as defined, for instance, in Refs.~\cite{Lay14,Kan13b}.

In relation to the Pauli exclusion principle,
if a method without blocking is used (WB),
the Hamiltonian in Eq.~(\ref{eq:H}) is directly the Hamiltonian of the system.
In this case, the Pauli principle is applied after the diagonalization of $\mathcal H$:
the obtained bound states that we consider forbidden by comparison with the spherical and Nilsson limits are discarded.
This is a usual way to deal with the Pauli principle;
however, the Pauli principle can have a strong effect on the calculation of energies and wave functions.
Therefore, in some cases, it is convenient to consider methods for blocking single-particle states occupied by the core nucleons.
In this work, we use the BCS theory to account for the partial occupation of the single-particle Nilsson states.

\subsection{\label{sec:Blocking}Implementation of BCS formalism}
\begin{figure}
\includegraphics[width=1.0\linewidth]{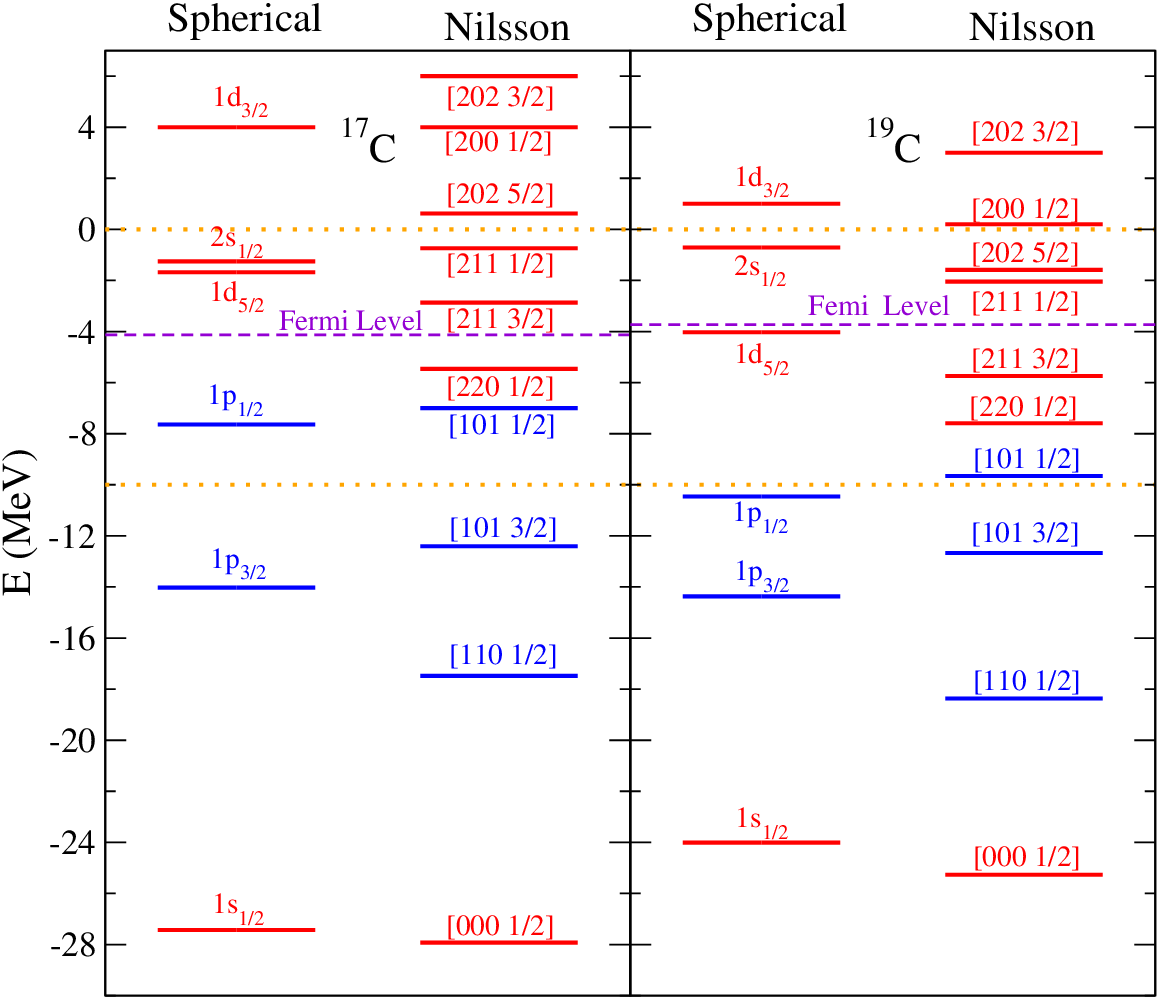}
\caption{\label{fig:nilsson_levels} Spherical single-particle levels and Nilsson levels of $^{17}$C (left) and $^{19}$C (right) obtained using the NAMD (PB) model.
The Fermi level obtained in the BCS calculation is also represented.
Positive and negative parity states are represented with different colours, and Nilsson levels are labelled with the asymptotic quantum numbers [$Nn_3\Lambda\ \Omega$] referring to large prolate deformations~\cite{BM}.}
\end{figure}

The Bardeen–Cooper–Schrieffer (BCS) formalism is applied as explained in Ref.~\cite{Brink_Broglia}. 
The eigenvalues $\varepsilon_\nu$ of the Nilsson Hamiltonian $\mathcal H_N=\mathcal H-h_\text{core}$ are taken as single-particle energies in the BCS calculation for the N neutrons of the core.
A constant pairing strength $G$ is considered 
to act between a set of levels around the Fermi energy $\lambda$. 
Levels from 0~MeV to 10~MeV below the neutron separation threshold are included in the BCS active space. The specific levels for each nucleus studied can be seen in Fig.~\ref{fig:nilsson_levels}, between the dotted lines
A new Hamiltonian can be defined in terms of the creation and annihilation operators~\cite{Brink_Broglia}
\begin{equation}
H_{BCS}=
\sum_{\nu}(\varepsilon_\nu-\lambda)(a^\dag_\nu a_\nu+a^\dag_{\bar\nu} a_{\bar\nu})
-G\sum_{\nu\nu'}a^\dag_\nu a^\dag_{\bar\nu}a_{\bar\nu'}a_{\nu'}.
\end{equation}
The single-particle energies $\varepsilon_\nu$ and their associated wave functions $\psi_\nu(\vec{r'})$ are obtained by diagonalizing the Hamiltonian $\mathcal H_N$ in the THO basis giving rise to the linear combinations,
\begin{equation}
\psi_\nu(\vec{r'})=
\sum_{nj}C^{\nu}_{nj} R^{THO}_{nl}(r)\mathcal Y^{j\Omega}_{\ell s}(\hat r'),
\end{equation}
where $C^{\nu}_{nj}$ are the components of the eigenvector matrix.
These Nilsson states $\nu$ do not have well-defined values of $\ell$ and $j$,
but they can be characterised by their parity $\pi$
and the projection $\Omega$ of $\vec j$ along the axial symmetry axis.
The energy $\varepsilon_\nu$ corresponds to the state $\nu$ with projection $\Omega$ and wave function $\psi_\nu(\vec{r'})$,
but also to the time-reversed state $\bar\nu$ with projection $-\Omega$ and wave function
\begin{equation}
\psi_{\bar\nu}(\vec{r'})=
\sum_j(-1)^{j-\Omega}C^{\nu}_{nj} R^{THO}_{nl}(r)\mathcal Y^{j-\Omega}_{\ell s}(\hat r').
\end{equation}
Therefore, unlike spherical single-particle levels,
all the Nilsson levels can be occupied only by a pair of neutrons or protons.
The schemes of spherical and Nilsson single-particle levels obtained for $^{17}$C and $^{19}$C are shown in Fig.~\ref{fig:nilsson_levels},  with the model parameters specified below.

The application of the BCS formalism involves the Bogoljubov-Valatin canonical transformation from
particles to quasiparticles.
Therefore, once the BCS calculation is performed,
the particle state $|\nu\rangle$ becomes a one-quasiparticle state $|\text{BCS}, \nu JM\rangle$.
Additionally, the state has been projected in the fixed laboratory system by including the rotational state of the nucleus. Then, for a total angular momentum $J$ with projection $M$ on the $Z$-axis of the laboratory system, the one-body wave function becomes
\begin{eqnarray}
\Psi_{\nu J}^M(\vec{r'},\omega)=
\frac{\sqrt{2J+1}}{4\pi}\left[
\psi_\nu(\vec{r'})
{\cal D}_{M\Omega}^{J}(\omega)^*\right.+\nonumber\\
\left. (-1)^{J-\Omega}
\psi_{\bar\nu}(\vec{r'}){\cal D}_{M-\Omega}^{J}(\omega)^*\right].
\label{eq:IntAng}\end{eqnarray}
The definition of \cite{BS} is used for the
rotation matrices ${\cal D}_{M\Omega}^{J}(\omega)$
and the three Euler angles are denoted by $\omega$.
The functions $\Psi_{\nu J}^M(\vec{r'},\omega)$
are orthonormal and take into account the symmetry 
regarding the $\Omega$ and $-\Omega$ projections.
The effect of $h_{\text{core}}$ is included by diagonalizing a new quasiparticle Hamiltonian
$H_{J^\pi}=H_{BCS}+h_{\text{core}}-e_0$ using states $|\text{BCS}, \nu JM\rangle$ as a basis.
$e_0\equiv\langle \text{BCS}|H_{BCS}+h_{core}|\text{BCS}\rangle$ is the energy of the BCS ground state, that is the quasiparticle vacuum.
The matrix elements of $H_{J^\pi}$ result:
\begin{eqnarray}
 \langle \text{BCS},\nu'JM|H_{J^\pi}|\text{BCS},\nu JM\rangle=\nonumber\\
 E_\nu\delta_{\nu'\nu}+\frac{\hbar^2}{2\mathcal J}
 \langle\Psi^M_{\nu' J}|\vec I^2|\Psi^M_{\nu J}\rangle\mathcal F_{\nu'\nu}.
\end{eqnarray}
$E_\nu$ is the quasiparticle energy associated with $\epsilon_\nu$ and the $\mathcal F_{\nu'\nu}$ factor
appears naturally when including a one-body operator between one-quasiparticle states~\cite{Brink_Broglia,BM}. 
It depends on the occupation numbers $u_\nu$ and $v_\nu$ obtained in the BCS calculation as 
\begin{equation}
\mathcal F_{\nu'\nu}=u_{\nu'}u_\nu-v_{\nu'}v_\nu.
\end{equation}

For each angular momentum and parity $J^\pi$, a set of energies $E_i^{J^\pi}$ with their corresponding eigenvectors having components $C^{iJ}_{\nu}$
are obtained.
The associated wave functions can be written in the fixed laboratory frame as in Eq.~(\ref{eq:LabWF}), 
with radial functions $R^{J^\pi}_{i\alpha}(r)=\sum_nC_{n\alpha}^{iJ^\pi}R^{THO}_{n\ell}(r)$.
The coefficients can be calculated from the eigenstates of $H_{J^\pi}$ and $\mathcal H_{N}$
\begin{equation}
C^{iJ^\pi}_{n\alpha}=
\sqrt{\frac{2I+1}{2J+1}}\sqrt{1+(-1)^I}
\sum_\nu\langle j\Omega I 0|J\Omega\rangle
C^{iJ}_{\nu }C^\nu_{nj},
\end{equation}
using expression~(8) from Ref.~\cite{Punta23}. We can also define the occupation number $v_{iJ^\pi}^2=\sum_\nu (C_\nu^{iJ}v_\nu)^2$ related to the occupation of the state by the core neutrons.

The obtained quasiparticle states mix particle and hole states.
However, we assume that the states with low occupation by core neutrons ($v_{iJ^\pi}^2<0.5$) can be approximated as particle states, while those with high occupation ($v_{iJ^\pi}^2>0.5$) are considered hole states and they are discarded.
The particle states have an associated energy $\varepsilon_i^{J^\pi}=E_i^{J^\pi}+\lambda$,
where $\lambda$ is the Fermi energy obtained in the BCS calculation.
These are  the energies  that can be compared with the eigenvalues of the Hamiltonian without blocking (\ref{eq:H}) and the experimental data.

Note that the factor $\mathcal F_{\nu'\nu}$ reduces the coupling between Nilsson states with high and low occupancy.
Therefore, this formalism can be regarded as a partial blocking method (PB).
It is also interesting to explore the extreme case of a total blocking method (TB).
To do this, we assume that the pairing strength is zero and the lowest N/2 Nilsson states are considered fully occupied ($u_\nu=0,\ v_\nu=1$), while the rest are completely empty ($u_\nu=1,\ v_\nu=0$).
In this case, the occupied and empty Nilsson states are completely decoupled because $\mathcal F_{\nu'\nu}=0$ between them.
This TB method is equivalent in practice to model III (``deformed PF model'') from Ref.~\cite{Shin24}.

\section{\label{sec:17C}Application to  $^{\textbf{17}}$C}
The NAMD model has been applied to study the $^{17}$C system using the three different blocking methods.
A slight renormalisation of the central potential $V_0(r)$  was necessary to match the energy of the ground state with the experimental value.
This renormalisation is different depending on the blocking model: 1.014 (WB), 1.000 (TB) and 0.997 (PB).
The spin-orbit potential is obtained as a function of the derivative of $V_0(r)$ following the relation from~\cite{Hamamoto05}, but using our semi-microscopic function instead of a Wood-Saxon potential.
For all blocking methods,
the value $\hbar/2\mathcal{J}=0.295~\text{MeV}$ is used for the core Hamiltonian,
compatible with the excitation energy of the first excited state $2^+$ of $^{16}\text{C}$ ($1.766~\text{MeV}$ \cite{TILLEY93}).
Regarding to the deformation length, $\delta=1.27$~fm (prolate) is obtained following the prescription of \cite{Lay14}.

In the PB method, a pairing strength $G=1.3$~MeV is considered between the four bound Nilsson states with separation energy between -10~MeV and the threshold.
More deeply bound states are considered fully occupied and unbound states are considered completely empty.
In the PB method, the first three deeply bound Nilsson states are considered fully occupied,
From the BCS calculation, the Fermi level is found at -4.13~MeV and the pairing gap is 1.58~MeV, close to the value obtained with the three-point formula centered on the $^{16}$C system, 1.76~MeV.
\begin{figure}
\includegraphics[width=1.0\linewidth]{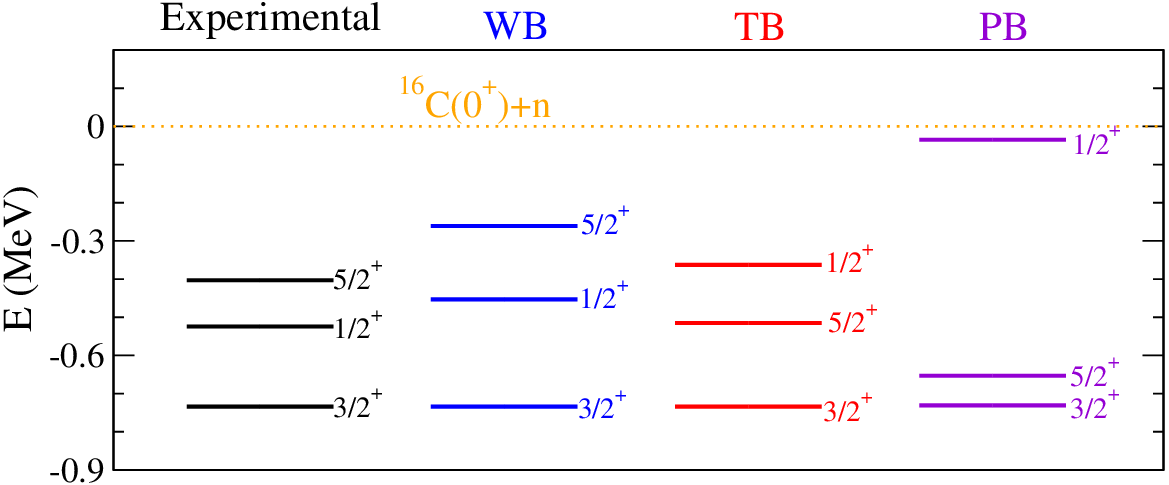}
\caption{\label{fig:levels_c17} Energies of the bound states of $^{17}$C.
The experimental values from~\cite{Wang_2017,Ele05} are compared with the results of the NAMD model using different blocking methods.}
\end{figure}

The Hamiltonian is diagonalized in the THO basis using the parameters $b=2.4\text{ fm}$ and $\gamma=3.0\text{ fm}^{1/2}$ and considering $0\leq\ell\leq4$ and $1\leq n\leq30$.
The energies of the bound states obtained for the different blocking methods are shown in Fig.~\ref{fig:levels_c17} compared to the experimental values~\cite{Wang_2017,Ele05}.
Using the TB and PB methods, the $5/2^+$ state becomes the first excited state and the $1/2^+$ the second, contrary to the experimental evidence.
However, it should be noted that the three bound states are very close in energy,
so that the difference between the calculated levels and the experimental values is less than 0.5~MeV.

\begin{figure}
\includegraphics[width=1.0\linewidth]{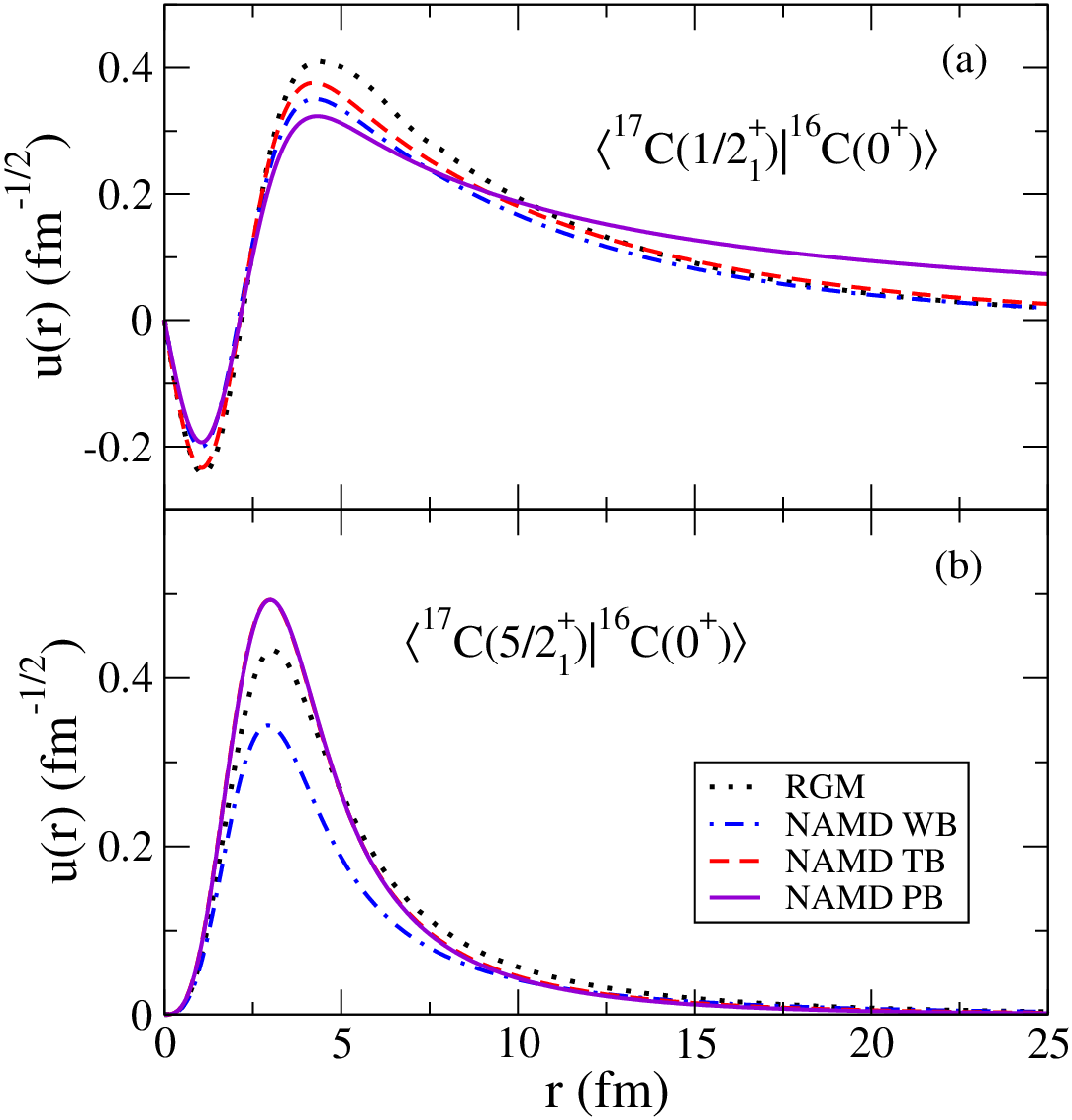}
\caption{\label{fig:wf_ex} Overlaps of the ground state of $^{16}$C with the first (upper panel) and second (lower panel) excited states of $^{17}$C.
The results of the RGM calculations~\cite{Chien23} are compared with the results of the NAMD model using different treatments of Pauli blocking.}
\end{figure}

To compare the wave functions obtained with each blocking method,
the radial overlap functions between the excited bound states of $^{17}$C and the ground state of the core $^{16}\text{C}(0^+)$ are shown in Fig.~\ref{fig:wf_ex}.
These are the relevant overlaps for the study of the $^{16}$C(d,p)$^{17}$C reaction that will be discussed later.
The overlaps obtained from the resonating group method (RGM)~\cite{Chien23}, are also shown.
The RGM method is a microscopic cluster model that is much more complex than the models presented here. For example, 990 Slater determinants are used to obtain the overlaps.
Therefore, it is natural to find differences between the NAMD and RGM models.
However, in the lower panel of Fig.~\ref{fig:wf_ex} it can be seen how,
by applying total or partial blocking,
the NAMD model produces overlaps $\langle^{17}\text{C}(5/2^+_1)|^{16}\text{C}(0^+)\rangle$ close to microscopic RGM calculation at large distances.
Note that transfer reactions are peripheral, and hence mostly sensitive to the behaviour of the nucleons around the surface of the nucleus.
On the other hand,  $\langle^{17}\text{C}(1/2^+_1)|^{16}\text{C}(0^+)\rangle$ are similar except for the PB case, where the mismatch in energy of the state leads to differences in the asymptotic behaviour.
\subsection{\label{sec:c16dpc17} $^{\textbf{16}}\textbf{C}\bm{(d,p)}^{\textbf{17}}\textbf{C}$}
\begin{figure}
\includegraphics[width=1.0\linewidth]{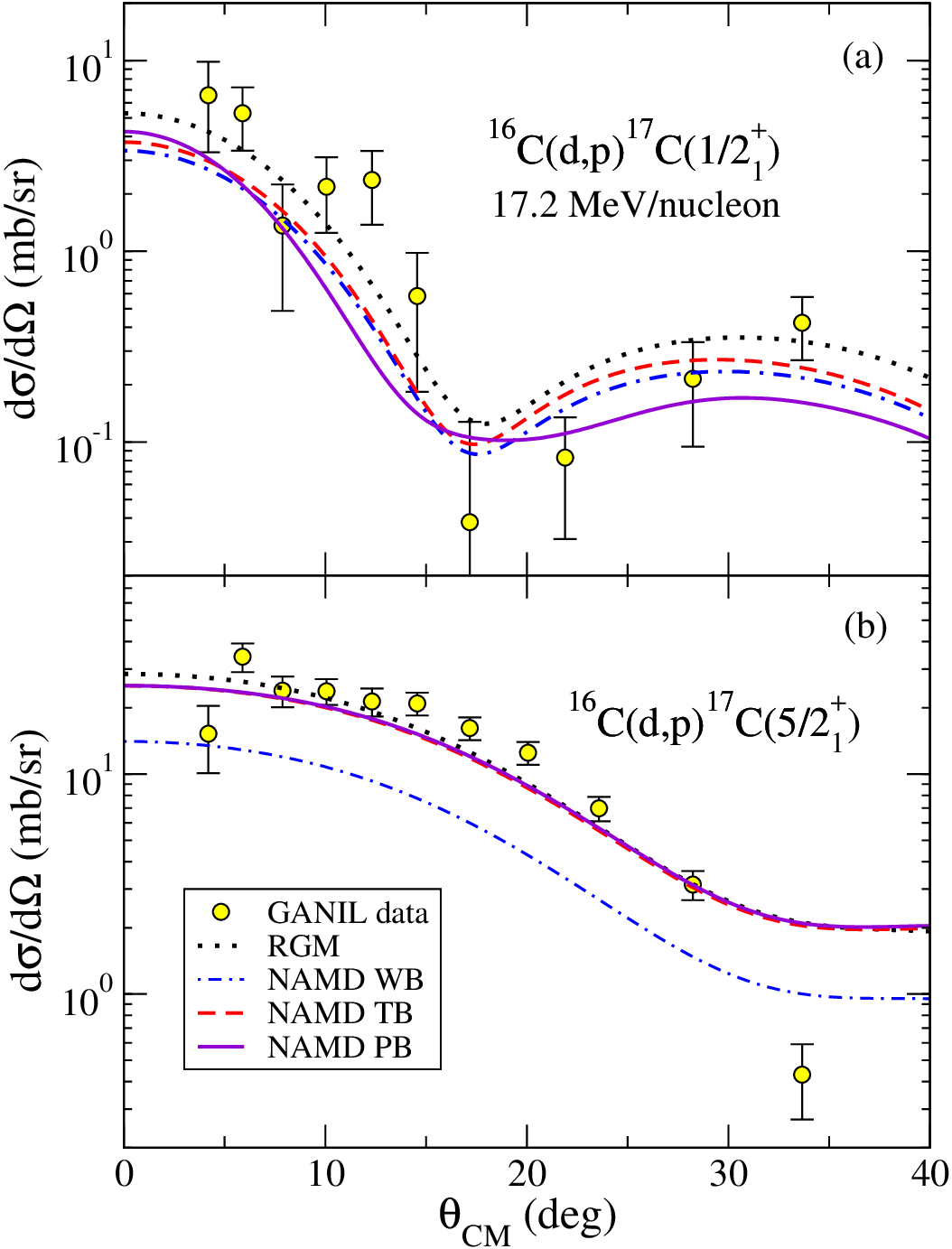}
\caption{\label{fig:c16dpc17_ex}  
Angular distribution of the $^{16}\text{C}(d,p)^{17}\text{C}$
reaction at 17.2 MeV/nucleon
when the $^{17}\text{C}$ bound states $1/2^+_1$ (upper panel),
and $5/2^+_1$ (lower panel) are populated.
The results using the different blocking methods are compared
with the experimental data \cite{Pereira} and the results using the RGM overlaps~\cite{Chien23}.}
\end{figure}

The differential cross sections of the transfer reaction $^{16}\text{C}(d,p)^{17}\text{C}$ populating bound excited states
have been calculated as in Ref.~\cite{Punta23}, using the finite-range adiabatic distorted wave approximation (ADWA) \cite{JT}, and employing the same optical model potentials for consistency.
In the \emph{post} form,
these calculations require the overlap functions $\langle ^{17}\text{C}|^{16}\text{C}(0^+)\rangle$, 
which are taken from the results of the NAMD model.
The results obtained using the three different blocking methods are compared with the experimental data from GANIL~\cite{Pereira} in Fig.~\ref{fig:c16dpc17_ex},
These data were obtained in inverse kinematics with a $^{16}$C beam at $17.2$~MeV/nucleon.

For the case of the first excited state $1/2^+$,
despite the difference of the PB due to its energy mismatch,
we find good agreement between the results for all blocking methods and the data (Fig.~\ref{fig:c16dpc17_ex}(a)).
The results without blocking (WB) are similar to those obtained in~\cite{Punta23},
which underestimate the differential cross section for the second excited state $5/2^+$ (Fig.~\ref{fig:c16dpc17_ex}(b)).
However, using the total and partial blocking methods (TB and PB),
 the agreement with the experimental data for this state is significantly improved.
Figure ~\ref{fig:c16dpc17_ex} also shows the ADWA results using microscopic RGM overlaps.
The agreement of the results using this RGM model with the data is rather good,
but it should be noted that, using a simpler model such as NAMD
a similar degree of agreement is found, provided that Pauli blocking effects are properly accounted for. 

\begin{table}
\caption{\label{tab:S0}%
Spectroscopic factors $S(0^+)$ of the bound states of \textsuperscript{17}C.
The results of the NAMD and RGM~\cite{Chien23} models are compared with those extracted from the experimental analysis~\cite{Pereira}.}
\begin{ruledtabular}
\begin{tabular}{ccccccc}
State&$(\ell,\ j)$&WB&TB&PB&RGM&Exp.\\
\hline
		$3/2^+_1$ & (2, 3/2) & 0.00 & 0.01 & 0.01 & 0.01 & 0.03 
  \begin{tabular}{@{}c@{}}+0.05 \\ -0.03\end{tabular}\\
		$1/2^+_1$ & (0, 1/2) & 0.68 & 0.80 & 0.82 & 0.94 & 0.64$\pm$0.18 \\
		$5/2^+_1$ & (2, 5/2) & 0.33 & 0.66 & 0.65 & 0.56 & 0.62$\pm$0.13
\end{tabular}
\end{ruledtabular}
\end{table}
In Table~\ref{tab:S0}, the spectroscopy factors (SF) extracted from the RGM calculations~\cite{Chien23} and the different NAMD models are compared with the experimental ones from Ref.~\cite{Pereira}.
These experimental SF are obtained as the ratio of the experimental cross sections and pure single-particle finite-rage ADWA calculations using CH89 parameterisation.
In Table~\ref{tab:S0}, it can be seen how the discrepancy of the NAMD WB model in $^{16}\text{C}(d,p)^{17}\text{C}(5/2^+_1)$ is related to the difference in SF.
It is also shown that there is agreement in the small contribution of the component $d_{3/2}\otimes0^+$ in the ground state of $^{17}$C.
This makes the $^{16}\text{C}(d,p)^{17}\text{C}$ cross section to this state small and difficult to measure experimentally;
to study the ground state of $^{17}\text{C}$ one may consider the reverse reaction $^{17}\text{C}(p,d)^{16}\text{C}$, which is studied in the next subsection.
\subsection{\label{sec:c17pdc16} 
$^{\textbf{17}}\textbf{C}\bm{(p,d)}^{\textbf{16}}\textbf{C}$}
The transfer reaction $^{17}\text{C}(p,d)^{16}\text{C}$ is studied at 31~MeV/nucleon so that the relative velocities $p-^{17}$C and $d-^{16}$C are the same as those of the previous section.
Consequently, using the same potentials and the overlap functions $\langle ^{17}\text{C}(3/2^+_1)|^{16}\text{C}\rangle$,
differential cross sections can be calculated using the ADWA approximation in \emph{prior} form.

\begin{figure}
\includegraphics[width=1.0\linewidth]{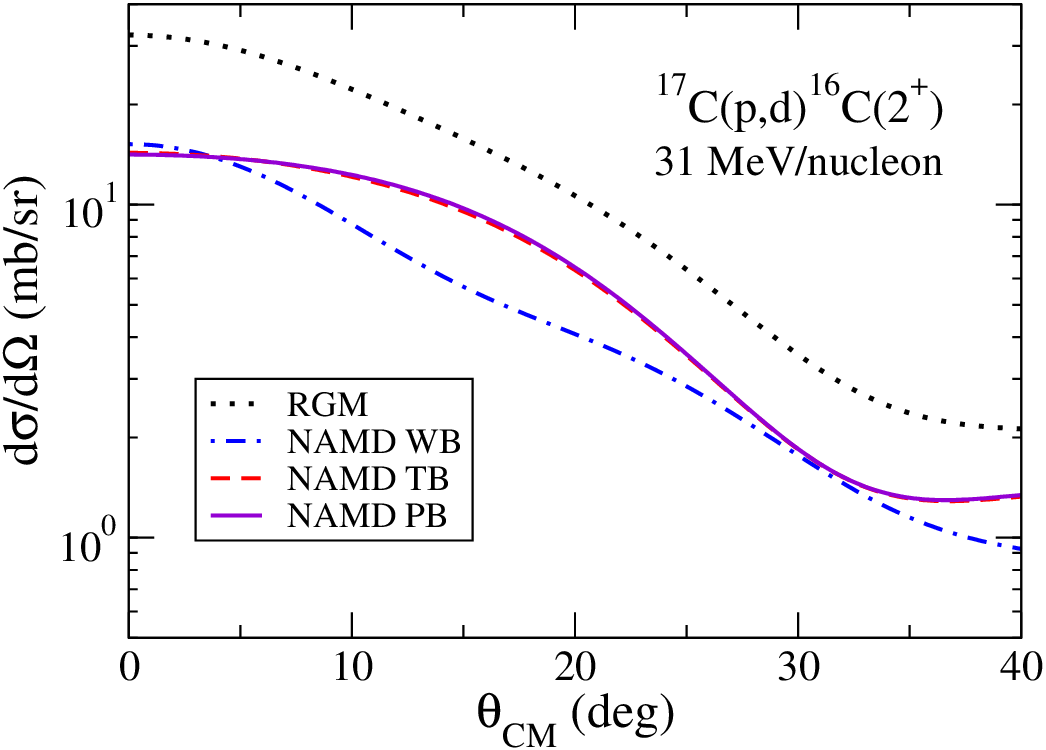}
\caption{\label{fig:c17pdc16}  
Differential cross section of $^{17}\text{C}(p,d)^{16}\text{C}$ transfer to the first excited state $2^+$ of $^{16}\text{C}$ at 31~MeV/nucleon.
The results using the NAMD model with different blocking methods are compared
with the results using the RGM overlaps~\cite{Chien23}.}
\end{figure}

The transfer to the ground state in $^{17}\text{C}(p,d)^{16}\text{C}$ is analogous to that in $^{16}\text{C}(d,p)^{17}\text{C}$.
In particular, both depend on the overlap function $\langle ^{17}\text{C}(3/2^+_1)|^{16}\text{C}(0^+)\rangle$.
Again, the small weight of this component makes the transfer cross section to the $0^+$ ground state  of $^{16}$C small.
The cross section of the transfer to the first excited state $^{16}\text{C}(2^+)$ must be larger and is expected to provide more information.
The results obtained in this case using the NAMD and RGM models are shown in Fig.~\ref{fig:c17pdc16}.
The differences between the models can be understood by looking at their corresponding SF, which are presented in Table~\ref{tab:S2}. 
On the one hand, the cross section is larger using the overlaps from the RGM calculations because its SF exceeds unity,
while the approximate values obtained with NAMD models are necessarily less than one.
On the other hand, the larger weight of the $s_{1/2}\otimes2^+$ component ($\ell=0$) in the RGM and NAMD WB models causes a difference in the shape of the distribution compared to the others,
which is the result of the clear dominance of the $\ell=2$ components.

\begin{table}
\caption{\label{tab:S2}%
Comparison of the spectroscopic factors $S(2^+)$ of the ground state of $^{17}$C obtained with the NAMD and RGM~\cite{Chien23} models.
The results for the component with $\ell=0$ and the sum of the two components with $\ell=2$ are shown.
The component with $\ell=4$ is lower than 0.001 in all cases.}
\begin{ruledtabular}
\begin{tabular}{ccccc}
$\ell$ &WB&TB&PB&RGM\\
\hline
		$0$ & 0.33 & 0.04 & 0.02 & 0.37 \\
		$2$ & 0.50 & 0.88 & 0.88 & 1.13
\end{tabular}
\end{ruledtabular}
\end{table}

\section{\label{sec:19C}Application to  $^{\textbf{19}}$C}
The NAMD model has also been applied to the study the $^{19}$C nucleus, using the same THO basis as in the case of $^{17}$C.
The spin-orbit potential is parametrised in terms of the derivative of the Wood-Saxon function with standard values of radius ($R_{so}$=3~fm), diffuseness ($a_{so}$=0.7~fm) and strength ($V_{so}$=6.5~MeV).
The value $\hbar/2\mathcal{J}=0.25\text{ MeV}$ is used for the core Hamiltonian,
compatible with the excitation energy of the first excited state $2^+$ of $^{18}\text{C}$ (1.588~MeV~\cite{Sta03}).
The deformation length of the NAMD model is that of the AMD calculation,
$\delta_2=1.20$~fm~\cite{Lay14} (prolate).
For the PB method, as in  $^{17}$C,
three deeply bound Nilsson states are considered fully occupied,
and the unbound states are considered completely empty.
The BCS calculation is performed with the five bound states between -10~MeV and the threshold.
Using $G=1.3$~MeV, 
the Fermi level is found at -3.73~MeV and the pairing gap is 1.78~MeV, compatible with the value obtained with the three-point formula centered on the $^{18}$C system, 1.80~MeV.

\begin{figure}
\includegraphics[width=1.0\linewidth]{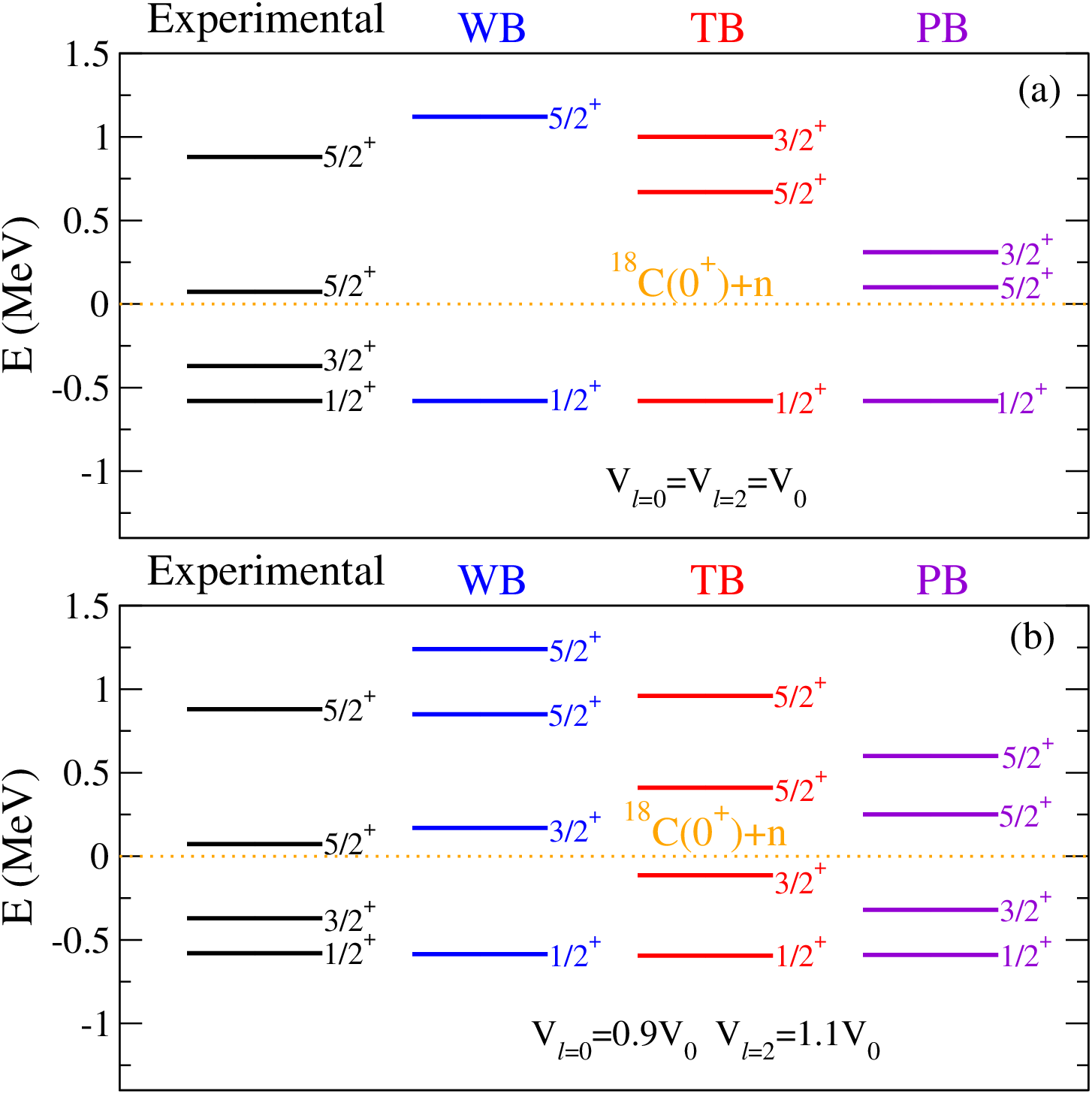}
\caption{\label{fig:levels_c19} Spectrum obtained for the $^{19}$C nucleus using the NAMD model considering different Pauli blocking methods in comparison with the experimental one~\cite{Ele05,Sat08,Thoen13,Hwang17}.
The upper panel shows the energies obtained without $\ell$-dependent renormalization,
while the lower panel shows the result after including this renormalization (see text).}
\end{figure}

In principle, the NAMD model gives almost degenerate spherical single-particle levels $2s_{1/2}$ and $1d_{5/2}$ around -2~MeV, and the $1d_{3/2}$ around 3~MeV. 
With this $sd$-shell scheme, $3/2^+$ and $5/2^+$ states of $^{19}$C are not found up to 1~MeV above the $1/2^+$ ground state, as shown in Fig.~\ref{fig:levels_c19}.(a).
In order to obtain a better agreement with the experimental spectrum~\cite{Ele05,Sat08,Thoen13,Hwang17},
a $\ell$-dependent renormalisation of the central potential $V_0(r)$ is applied:
0.9 for $\ell=0$ and 1.1 for $\ell=2$.
Thus, making the $1d$ levels more bound with respect to the $1s$ level,
the $3/2^+$ and $5/2^+$ states of $^{19}$C are closer to the ground state $1/2^+$.
Furthermore, the potential $V_0(r)$ continues to be slightly renormalised globally in each model to match the energy of the ground state with the experimental value: 0.973~(WB), 0.972~(TB), 0.998~(PB).

In Fig.~\ref{fig:levels_c19}(b), it can be seen that in this way the spectra obtained with the TB and PB models are close to the experimental one, whereas significant deviations are seen in the case of the WB model.
Note that we are assuming that the first $5/2^+$ state is a low-energy resonance, as suggested in~\cite{Thoen13,Hwang17}, rather than a bound state as previously reported~\cite{Ele05}.

\subsection{\label{sec:c18dpc19} $^{\textbf{18}}\textbf{C}\bm{(d,p)}^{\textbf{19}}\textbf{C}$}
The wavefunctions obtained with the NAMD models are applied to the calculation of the transfer reaction $^{18}\text{C}(d,p)^{19}\text{C}$ at 17~MeV/nucleon.
At this energy, the $d-^{18}$C system has a relative energy very similar to that of the $d-^{16}$C system in the previous reactions.

\begin{figure}
\includegraphics[width=1.0\linewidth]{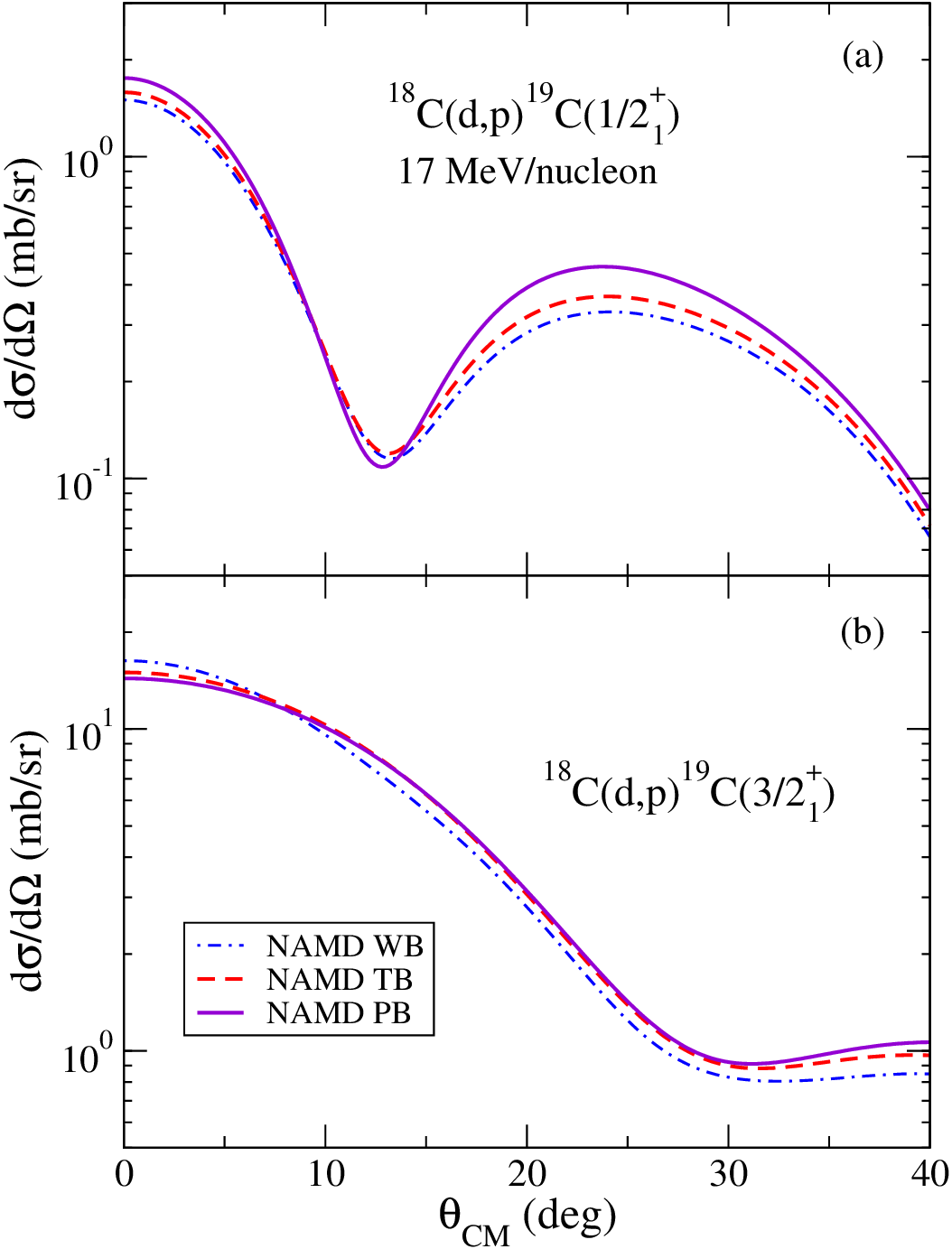}
\caption{\label{fig:c18dpc19}  
Angular distribution of the $^{18}\text{C}(d,p)^{19}\text{C}$ reaction 
when the $^{19}\text{C}$ ground state $1/2^+_1$ (upper panel),
and first excited state $3/2^+_1$ (lower panel) are populated.
Calculations have been made for 17~MeV/nucleon using the NAMD model with different blocking methods.}
\end{figure}

The transfer to the ground state and the first excited state of $^{19}$C are studied and the differential cross sections obtained are shown in Fig.~\ref{fig:c18dpc19}.
The results of the models are not very different in this case,
which can be explained by comparing the calculated SF.
The SF are shown in Table~\ref{tab:S0_2}, and it can be seen that the values are very close,
approximately 0.7 for the $s_{1/2}\otimes0^+$ component in the ground state and 0.5 for the $d_{3/2}\otimes0^+$ component in the first excited state, pointing to a smaller importance of Pauli blocking effects in this case.

\begin{table}
\caption{\label{tab:S0_2}%
Spectroscopic factors $S(0^+)$ of the bound states of \textsuperscript{19}C.
The results obtained using different Pauli blocking method in the NAMD model are compared.}
\begin{ruledtabular}
\begin{tabular}{ccccc}
State&$(\ell,\ j)$&WB&TB&PB\\
\hline
		$1/2^+_1$ & (0, 1/2) & 0.68 & 0.72 & 0.69 \\
		$3/2^+_1$ & (2, 3/2) & 0.52 & 0.52 & 0.54
\end{tabular}
\end{ruledtabular}
\end{table}

\section{\label{sec:Conclusions}Summary and Conclusions}
A new two-body model based on the combination of Nilsson and PAMD models from Ref.~\cite{Punta23},
has been applied to the study of weakly bound exotic nuclei.
The model considers a neutron moving in a deformed potential generated by the core.
The interaction includes semi-microscopic coupling potentials calculated using AMD transition densities~\cite{Lay14}, a spin-orbit potential, and a collective rotational term.
By incorporating the BCS formalism in the model, different Pauli blocking methods are used,
and their results are compared.
Therefore, we have core+nucleon models suitable for application to the study of reactions, providing a very complete description of the nuclei, since they consider the deformation of the core including microscopic information, the collective rotation, and the pairing effects in case of PB.
The models have been applied to the description of $^{17}$C, $^{19}$C and transfer reactions that involve these nuclei: $^{16}\text{C}(d,p)^{17}\text{C}$, $^{17}\text{C}(p,d)^{16}\text{C}$, $^{18}\text{C}(d,p)^{19}\text{C}$.

Although all variants of the NAMD model give a reasonable description of $^{17}$C,
including Pauli blocking methods improves the agreement with the data for the stripping reaction $^{16}\text{C}(d,p)^{17}\text{C}(5/2^+_1)$.
Our results are compared with those obtained using microscope RGM calculations~\cite{Chien23},
which are in good agreement with the experimental data.
It should be noted that using a much simpler model such as NAMD, if we include blocking effects,
the agreement with the data is similar to that found with this microscopic model.
Blocking the Nilsson state $[220\ 1/2]$ (see Fig.~\ref{fig:nilsson_levels}) which is dominated by the $s_{1/2}$ component,
implies a reduction of the $s_{1/2}\otimes2^+$ strength in the second excited state $5/2^+_1$, 
increasing the weight of the component $d_{5/2}\otimes0^+$.
This is reflected in the SF shown in Table~\ref{tab:S0},
or in the complete $\langle^{17}\text{C}(5/2^+_1)|^{16}\text{C}(0^+)\rangle$ overlaps shown in Fig.~\ref{fig:wf_ex}.
The increase in such overlap explains therefore the better agreement with the experimental data of $^{16}\text{C}(d,p)^{17}\text{C}(5/2^+_1)$ as compared to the calculation in which Pauli blocking is not taken into account.
TB and PB methods give goods results in $^{16}\text{C}(d,p)^{17}\text{C}$ reaction, 
but they predict the spectrum of the bound states in a different order than that observed experimentally.
Notice, however, that differences in energies are very small.
The energy mismatch is worse in the case of PB, and this could be indicative that a better implementation of pairing would be needed.
Blocking methods also drastically reduce the $s_{1/2}$ strength  in the ground state,
and it can be seen in the transfer reaction $^{17}\text{C}(p,d)^{16}\text{C}(2^+)$.
Obtaining experimental data for this reaction would help determine whether the ground state of $^{17}$C is strongly dominated by $\ell=2$ components as suggested by the TB and PB models,
or if there is a significant contribution from $\ell=0$ as suggested by the WB and the microscopic RGM model. 

In case of $^{19}$C, an $\ell$-dependent renormalisation is needed to reasonably describe the system using the NAMD model.
This limitation of our model can be related to the assumption of a prolate deformation,
when the possibility that prolate and oblate structures are almost degenerate (shape co-existence) has been previously discussed~\cite{Suz03}.
Using the $\ell$-dependent renormalisation,
the NAMD TB model obtains a spectrum up to 1.5~MeV that agrees with the experimental one, 
the agreement not being so good in the case of the WB model.
Including pairing, the PB model even improves the agreement.
However, there are no major differences in the $\langle^{19}\text{C}|^{18}\text{C}(0^+)\rangle$ overlaps obtained with the different models,
as reflected in the SF and the cross section of the reaction $^{18}\text{C}(d,p)^{19}\text{C}$.
It should be noted that experimental measurements of this transfer reaction would help confirm or rule out the shape coexistence in $^{18}$C and $^{19}$C.

The results shown here prove the suitability of the NAMD model to describe the $^{17}$C and $^{19}$C systems.
Furthermore, it has been proven that blocking of Nilsson states occupied by the core nucleons has significant effects that improve the description of the nuclei.
Although the partial blocking method (PB) is more complete,
the limit of total blocking (TB) proves to be a good approximation in the cases studied.
The application of these models to transfer to the continuum and breakup reactions is in progress and its extension to other weakly bound nuclei is planned.

\begin{acknowledgments}

We are grateful to P. Descouvemont for providing us with the RGM overlaps of Ref.~\cite{Chien23} and to B. Fern\'andez-Dom\'inguez for providing us with the cross section data of Ref.~\cite{Pereira}.
The present research is funded from grants PID2020-114687GB-I00 and PID2023-146401NB-I00 by MICIU/AEI/10.13039/501100011033,
P.P. acknowledges PhD grant FPU21/03931 and mobility grant EST24/00108 from the
Ministerio de Ciencia, Inovaci\'on y Universidades.

\end{acknowledgments}



\bibliography{punta,c19p}

\end{document}